# Possible exotic neutrino signature in electron muon collisions

Jai Kumar Singhal

Department of Physics
Government College, Sawai Madhopur-322 001, India
Email address: *jksinghal@hotmail.com*, singhal_ph@sancharnet.in

Sardar Singh and A K Nagawat

Department of Physics, University of Rajasthan, Jaipur-302 004, India

## ABSTRACT

With the strong experimental evidence for standard neutrino mass and mixings, there exists now a possibility of the lepton flavor violating process $e^+\mu^- \to W^+W^-$, which would occur via *t*-channel neutrino exchange induced by neutrino mixings. We consider Langacker's generalized neutrino mixings including ordinary (canonical SU(2)$_L \times$U(1)$_Y$ assignments), exotic (non-canonical SU(2)$_L \times$U(1)$_Y$ assignments) and singlet neutrinos leading to light and heavy mass eigen states. Constraints on lepton flavor violating (LFV) ordinary and heavy neutrino overlap parameters are obtained by using the current experimental bounds on LFV process $\mu \to e\gamma$. These constraints are used to analyze the dependence of differential cross section and angular distribution, for the process $e^+\mu^- \to W^+W^-$, on the mass of heavy (exotic) neutrino and c. m. energy ($\sqrt{s}$). The possibility of obtaining signatures of exotic neutrino mixings at *e*-$\mu$ collider is discussed.

# I. INTRODUCTION

With the convincing experimental evidence that standard neutrinos must have non-zero masses and these states are mixed [1, 2] the understanding of the neutrino sector remains one of the current issues. In principle new possible neutral heavy states may exist. Neutral heavy leptons beyond the Standard Model (SM) content arise in many extensions of the Standard Model, such as grand unified theories or super string inspired models [3, 4]. They are referred to as exotic neutrinos if they do not have usual $SU(2)_L \times U(1)_Y$ quantum numbers. So far none of these new states are experimentally ascertained and their masses are bounded to be greater than 80.5 - 90.3 GeV [1]. If exotic neutrinos exist then even in the case in which these new states are too heavy to be produced, their presence could still manifest through their mixing with the SM neutrinos. The phenomenology of the new heavy neutrinos has been studied extensively [5, 6].

One possibility to explore the neutrino sector is to consider lepton flavor violating (LFV) processes of standard particles, which are absent in the SM and depend on neutrino mass and mixings. There have been numerous theoretical and experimental studies of the LFV processes such as $\mu \rightarrow e\gamma$, $\mu \rightarrow 3e$, $\tau \rightarrow 3\mu$, muon to electron conversion in nuclei and muonium to antimuonium conversion [1, 7 - 8]. Yet another possibility is to consider lepton – lepton scattering. This mode is phenomenological attractive due to different kinematical range and experimental conditions [8]. The electron-muon collision seems to an interesting avenue to verify the properties of new exotic heavy neutrinos [8-13]. Several advantages of the $e$-$\mu$ colliders were pointed out in Ref. [10, 11]. The main advantages of the *e*-μ colliders over $e^+e^-$ colliders are: (i) muons beams are well known to have a reduced synchrotron radiation loss and (ii) the absence of the Z-mediated s-channel makes cleaner the high-energy properties of the charged current interactions. An interesting point is the fact that *e*-μ collider will test directly the properties of the two leptonic families [12, 13]. In this paper, we consider the process $e^+\mu^- \to W^+W^-$, which violates the lepton flavor and could occur through neutrino exchange via *t*-channel Feynman diagrams when the light-heavy neutrino mixing is considered. The observation of the process $e^+\mu^- \to W^+W^-$ would be a possible signal of the existence of the exotic heavy neutrinos and their mixing with light neutrinos.

## II. NEUTRINO MIXING FORMALISM

A comprehensive analysis of the mixing between ordinary fermions with canonical $SU(2)_L \times U(1)_Y$ assignments (i.e., left handed (L) fermions as $SU(2)_L$ doublets and right handed (R) fermions as $SU(2)_L$ singlets) and possible heavy fermions with exotic (non-canonical) $SU(2)_L \times U(1)_Y$ assignments (i.e., L-fermions as $SU(2)_L$ singlets and R- fermions as $SU(2)_L$ doublets) has been performed by Langacker and London [4]. In this analysis new neutrinos $N_{1L}$, $N_{2L}$ and $N_{3L}$ in the representation

$$\begin{bmatrix} N_1 \\ E^- \end{bmatrix}_L, \quad \begin{bmatrix} E^+ \\ N_2 \end{bmatrix}_L, \quad N_{3L}, \tag{1}$$

are considered. In the general Majorana case, the all $N_{1L}$, $N_{2L}$ and $N_{3L}$ can mix with the standard neutrinos. In the Dirac case, only $N_{1L}$ and $N_{3L}$ can mix with the standard neutrinos. Following Ref. [4] we consider the three SU(2)$_L\times$U(1)$_Y$ assignments,

$$\begin{bmatrix} n_{OL}^0 \\ e_L^{0\,-} \end{bmatrix}, \quad \begin{bmatrix} e_L^{0\,+} \\ n_{EL}^0 \end{bmatrix}, \quad n_{SL}^0, \qquad (2)$$

where $e_L^{0\,-}$ and $e_L^{0\,+}$ are weak eigenstates of charged leptons ($e_L^-$, $\mu_L^-$, $\tau_L^-$,..........) and anti-leptons ($e_L^+$, $\mu_L^+$, $\tau_L^+$,..........) respectively. The $n_{OL}^0$ are ordinary SU(2) doublet neutrinos, $n_{EL}^0$ are exotic neutrinos, occurring in doublets with left handed anti-leptons and $n_{SL}^0$ are SU(2) singlets neutrinos. These are themselves column vectors consisting of individual neutrinos. Let their dimensions are $n$-, $m$- and $p$- respectively [6], i.e.,

$$n_{OL}^0 = (\nu_{eL}^0 \ \nu_{\mu L}^0 \ \nu_{\tau L}^0 \ \nu_{4L}^0 \ \ldots \ \nu_{nL}^0)^T, \qquad (3)$$

$$n_{EL}^0 = (N_{1L}^0 \ N_{2L}^0 \ \ldots \ N_{mL}^0)^T, \qquad (4)$$

$$n_{SL}^0 = (N_{(m+1)L}^0 \ N_{(m+2)L}^0 \ \ldots \ N_{(m+p)L}^0)^T. \qquad (5)$$

For convenience, all of the weak eigenstate neutrinos are arranged into a column vector [4]

$$n_L^0 = (n_{OL}^0 \quad n_{EL}^0 \quad n_{SL}^0)^T. \qquad (6)$$

Following Ref. [4] we assume that the mass eigenstate neutrinos are all either light (*massless* i.e. with masses too small to be kinematically relevant) or heavy allowing the arrangement of the mass eigenstate into a column vector,

$$n_L = (n_{lL} \quad n_{hL})^T. \qquad (7)$$

The $n_{lL}$ and $n_{hL}$ are themselves column vectors consisting of $q$ light and $r$

heavy neutrinos respectively, i.e.,

$$n_{lL} = (\nu_{1L} \quad \nu_{2L} \quad .......... \quad \nu_{qL})^T, \tag{8}$$

$$n_{hL} = (N_{1L} \quad N_{2L} \quad .......... \quad N_{rL})^T. \tag{9}$$

The weak and mass eigenstates are related by unitary transformation

$$n_L^0 = U_L \, n_L, \tag{10}$$

with
$$U_L = \begin{bmatrix} A_L & E_L \\ F_L & G_L \\ H_L & J_L \end{bmatrix}, \tag{11}$$

where $A_L$, $F_L$ and $H_L$ are $(n \times q)$-, $(m \times q)$- and $(p \times q)$-dimension matrices describing the overlap of light neutrinos ($n_{lL}$) with ordinary doublets ($n_{OL}^0$), exotic doublets ($n_{EL}^0$) and singlets ($n_{SL}^0$) respectively. Similarly $E_L$, $G_L$ and $J_L$ are $(n \times r)$-, $(m \times r)$- and $(p \times r)$-dimension matrices describing the overlap of heavy neutrinos ($n_{hL}$).

The leptonic charged current involving light charged leptons is

$$\frac{1}{2} J_W^{\mu\dagger} = \bar{n}_{OL}^0 \gamma^\mu e_L^0 + \bar{n}_{ER}^{0c} \gamma^\mu e_R^0, \tag{12}$$

where $n_{ER}^{0c}$ are the neutrinos occurring in exotic R-doublets and related by CP to $n_{ER}^0$ [4].

The charged current in mass eigenstate basis is obtained by using Eq. (10) in Eq. (12) and is given by

$$\frac{1}{2} J_W^{\mu\dagger} = \bar{n}_{lL} \gamma^\mu A_L^\dagger e_L + \bar{n}_{hL} \gamma^\mu E_L^\dagger e_L + \bar{n}_{lR}^c \gamma^\mu F_R^\dagger e_R + \bar{n}_{hR}^c \gamma^\mu G_R^\dagger e_R \tag{13}$$

The first term in the charged current Eq. (13) represents non-universal reduction in the strength of usual left-handed current due to light-heavy neutrino mixing by the factor $A_L^\dagger$, which differ from unity. The second term represents the induced left-handed charged current between light charged and neutral heavy (exotic) states (i.e. coupling between the light charged and neutral exotic heavy neutral mass eigen states are induced) and these terms are phenomenologically important as they determine the decays and production of heavy neutrinos. Third and fourth terms represent the right-handed current induced by mixing of light-heavy neutrinos.

## III. THE PROCESS $e^+\mu^- \to W^+W^-$

### A. AMPLITUDE AND CROSS SECTION

The process $\mu^-(k_1,\sigma) + e^+(\bar{k}_1,\bar{\sigma}) \to W^-(k_2,\lambda) + W^+(\bar{k}_2,\bar{\lambda})$ (where the arguments indicate the four momenta and helicities of the respective particles), with the inclusion of neutrino mixing, occurs via $q$-light neutrinos $\nu_{\alpha L}$ ($\alpha = 1$ to $q$) and $r$ heavy neutrinos $N_{\xi L}$ ($\xi = 1$ to $r$) exchange in $t$-channel Feynman diagrams [Fig. 1]. We evaluate the amplitude for the process following the technique discussed by Renard [14] and Hagiwara and Zeppenfeld [15]. We separate the contributions to the amplitude in following two parts:

$$M_{\sigma\bar{\sigma},\lambda\bar{\lambda}} = M^\nu_{\sigma\bar{\sigma},\lambda\bar{\lambda}} + M^N_{\sigma\bar{\sigma},\lambda\bar{\lambda}} \qquad (14)$$

The $M^\nu_{\sigma\bar{\sigma},\lambda\bar{\lambda}}$ is the contribution due to light-neutrino exchange, and is found to be

$$M^\nu_{\sigma\bar{\sigma},\lambda\bar{\lambda}} = \frac{\sqrt{2}g^2}{D_\nu} d^{J_0}_{\Delta\sigma,\Delta\lambda} \sum_{\alpha=1}^{q} (A_L)_{e\alpha} \left[ \begin{array}{c} -\sqrt{2}\delta_{J_0,2} \\ + \frac{D_\nu}{2\beta}\left(B_{\lambda\bar{\lambda}} - \frac{C_{\lambda\bar{\lambda}}}{D_\nu}\right)\delta_{J_0,1} \end{array} \right] (A_L^\dagger)_{\alpha\mu} \delta_{\Delta\sigma,-1} \qquad (15)$$

with
$$D_\nu = 1 + \beta^2 - 2\beta \cos\theta . \qquad (16)$$

Here $\sqrt{s}$ is the total c. m. energy, $\beta = \sqrt{1 - 4m_W^2/s}$, $\gamma = \frac{\sqrt{s}}{2m_W}$, $\theta$ is the scattering angle, $\Delta\lambda = \lambda - \bar{\lambda}$, $\Delta\sigma = (\sigma - \bar{\sigma})/2$ (lepton helicities are normalized to ±1), $J_0 = \max(|\Delta\sigma|, |\Delta\lambda|)$ is minimum angular momentum. The $d^{J_0}_{\Delta\sigma,\Delta\lambda}$ are the d functions (see Table 2 of Ref. (15)). The coefficients $B$ and $C$ are as those given in Table I.

The heavy neutrino exchange contribution $M^N_{\sigma\bar{\sigma},\lambda\bar{\lambda}}$ is found to be

$$M^N_{\sigma\bar{\sigma},\lambda\bar{\lambda}} = \frac{\sqrt{2}\,g^2}{D_N} d^{J_0}_{\Delta\sigma,\Delta\lambda} \sum_{\xi=1}^{r} (E_L)_{e\xi} \left[ \begin{array}{c} -\sqrt{2}\,\delta_{J_0,2} \\ +\frac{D_\nu}{2\beta}\left(B_{\lambda\bar{\lambda}} - \frac{C_{\lambda\bar{\lambda}}}{D_\nu}\right)\delta_{J_0,1} \end{array} \right] (E_L^\dagger)_{\xi\mu}\, \delta_{\Delta\sigma,-1} \qquad (17)$$

with
$$D_N = 1 + \beta^2 - 2\beta \cos\theta + \frac{4m_\xi^2}{s} . \qquad (18)$$

Here a large number of unknown masses ($m_\xi$) and mixing parameters ($A$, $E$) make a thorough analysis impractical. However, if we consider the case of mass degenerate heavy neutrinos [16] or only one heavy neutrino then the discussion become tractable. For these cases the amplitude (Eq. (14)) becomes

$$M_{\sigma\bar{\sigma},\lambda\bar{\lambda}} = \sqrt{2}\,g^2 d^{J_0}_{\Delta\sigma,\Delta\lambda} \left( \frac{(A_L A_L^\dagger)_{e\mu}}{D_\nu} + \frac{(E_L E_L^\dagger)_{e\mu}}{D_N} \right) \\ \times \left[ -\sqrt{2}\,\delta_{J_0,2} + \frac{D_\nu}{2\beta}\left(B_{\lambda\bar{\lambda}} - \frac{C_{\lambda\bar{\lambda}}}{D_\nu}\right)\delta_{J_0,1} \right] \delta_{\Delta\sigma,-1} . \qquad (19)$$

From the unitarity of the mixing matrix ($U_L$,) we have $\left(A_L A_L^\dagger\right)_{e\mu} = -\left(E_L E_L^\dagger\right)_{e\mu}$. Using this in Eq. (19), we get

$$M_{\sigma\bar{\sigma},\lambda\bar{\lambda}} = -\sqrt{2}\, g^2\, d^{J_0}_{\Delta\sigma,\Delta\lambda} \left(\frac{4m_\xi^2/s}{D_\nu D_N}\right)\left(E_L E_L^\dagger\right)_{e\mu}$$
$$\times \left[-\sqrt{2}\delta_{J_0,2} + \frac{D_\nu}{2\beta}\left(B_{\lambda\bar{\lambda}} - \frac{C_{\lambda\bar{\lambda}}}{D_\nu}\right)\delta_{J_0,1}\right]\delta_{\Delta\sigma,-1} \quad (20)$$

With unpolarized initial beams the differential cross section for the production of unpolarized $W^+W^-$ pair is expressed as

$$\frac{d\sigma}{d\cos\theta} = \frac{\beta}{128\pi s} \sum_{\sigma\bar{\sigma},\lambda\bar{\lambda}} \left|M_{\sigma\bar{\sigma},\lambda\bar{\lambda}}\right|^2 . \quad (21)$$

The detailed expression of differential cross section is found to be

$$\frac{d\sigma}{d\cos\theta} = \frac{g^4}{128\pi\beta s} \frac{m_\xi^4}{m_W^4} \frac{1}{D_\nu^2 D_N^2} \left|\left(E_L E_L^\dagger\right)_{e\mu}\right|^2$$
$$\times \left(\begin{array}{c} \sin^2\theta\left\{\dfrac{2\beta^2}{\gamma^4}(D_\nu + 2\cos^2\theta) + \left(D_\nu - \dfrac{1}{\gamma^4}\right)^2\right\} \\ + \dfrac{(1-\cos\theta)^2}{\gamma^2}\left(D_\nu - \dfrac{1+\beta}{\gamma^2}\right)^2 + \dfrac{(1+\cos\theta)^2}{\gamma^2}\left(D_\nu - \dfrac{1-\beta}{\gamma^2}\right)^2 \end{array}\right) \quad (22)$$

Here it is appropriate to note that if instead of unpolarized initial ($e^+\mu^-$) beams, we take left-handed $\mu^-$ and right handed $e^+$ beams, the cross section is four times the unpolarized case [neutrinos couples only to the $\mu_L^-(e_R^+)$].

## B. ESTIMATION OF MIXING PARAMETER $\left|\left(E_L E_L^\dagger\right)_{e\mu}\right|^2$

For estimation of the mixing parameter $\left|\left(E_L E_L^\dagger\right)_{e\mu}\right|^2$ we use the current experimental bound on the branching ratio (BR) of the LFV process $\mu \to e\gamma$. The current experimental limit on the BR of the $\mu \to e\gamma$ is $1.2 \times 10^{-11}$ [17]. Langacker and London [18] discussed the neutrino mixing aspect and its effect on the BR for the process μ → e γ with the inclusion of light-heavy neutrino mixing. For the specific case considered here the BR (μ → eγ) [19] is (see Eq. (35) of Ref. [6])

$$BR(\mu \to e\gamma) = \frac{3\alpha}{32\pi}[F(x)]^2 |\lambda_{e\mu}^L|^2,  \quad (23)$$

where

$$|\lambda_{e\mu}^L|^2 = |(E_L E_L^\dagger)_{e\mu}|^2, \quad (24)$$

and

$$F(x) = \frac{x(1 - 6x + 3x^2 + 2x^3 - 6x^2 \ln x)}{(1-x)^4}, \quad (25)$$

with

$$x = m_\xi^2 / m_W^2.$$

The $|\lambda_{e\mu}^L|^2$ is constrained by the present experimental limit BR (μ → eγ) as a function of heavy neutrino mass $m_\xi$. These constraints are shown in Fig. 2.

## C. NUMERICAL RESULTS

We now present our numerical analysis of heavy neutrino mixing effects in the process $e^+ \mu^- \to W^+ W^-$. For numerical evaluation the values of mixing parameter $|(E_L E_L^\dagger)_{e\mu}|^2$ are taken as those obtained by Eq. (23) from the experimental constraints of the process μ → eγ.

In Fig. 3, the differential cross section $\frac{d\sigma}{d\cos\theta}$ is shown as a function of heavy neutrino mass ($m_\xi$) for different c. m. energies $\sqrt{s}$ = 200, 500, 800 and 1000 GeV at θ = $45^0$. We note that at lower center of mass energies (near production threshold for $W^+W^-$ pair) the differential cross section decreases with increase in heavy neutrino mass, whereas this trend is not observed at higher c. m. energies.

In Fig. 4, we plot the differential cross section $\frac{d\sigma}{d\cos\theta}$ as a function of c. m. energy $\sqrt{s}$ for varying heavy neutrino mass ($m_\xi$) from 100 to 500 GeV at $\theta = 45^0$. It is observed that:

(i) For a given heavy neutrino mass, the differential cross section first increases with increase in $\sqrt{s}$, attains a maximum and then decreases with increase in $\sqrt{s}$.

(ii) At lower c. m. energies ($\sqrt{s}$ < 200 GeV) for a given $\sqrt{s}$ the differential cross section decreases with increase in heavy neutrino mass ($m_\xi$), whereas this trend is reversed at higher c. m. energies ($\sqrt{s}$ > 300 GeV) for a fixed $\sqrt{s}$.

The angular distribution is displayed in Fig. 5, for different heavy neutrino mass $m_\xi$ = 100, 200 and 500 GeV at fixed $\sqrt{s}$ = 1000 GeV. We note that the angular distribution is enhanced with increase in $m_\xi$. For small heavy neutrino mass ($m_\xi \leq$ 200 GeV), the distribution is peaked at $\cos\theta \approx 1$.

## IV. CONCLUSIONS

The process $e^+\mu^- \to W^+W^-$ violates the lepton flavor and would occur if ordinary – exotic neutrino weak states are allowed to mix. Constraints on light (ordinary) –heavy (exotic) neutrino mixings are obtained by using experimental bounds on LFV process $\mu \to e\gamma$. Stringent constraints (|mixing parameter|$^2$

$\leq 10^{-8}$ ) are obtained for exotic neutrino mass $m_\xi \geq 5$ TeV (Fig. 2). For $m_\xi < 5$ TeV the constraints are less severe.

The amplitude for the process $e^+ \mu^- \to W^+ W^-$ depends on exotic heavy neutrino mass $m_\xi$ and reduces to zero in the absence of exotic neutrino mixings. The differential cross section (Fig. 3 and 4) and angular distribution (Fig. 5) depend on the heavy neutrino mass $m_\xi$. As such, measurements of these quantities would in principal give an estimate of the exotic neutrino mass $m_\xi$.

## ACKNOWLEDGEMENT

The authors are grateful to University Grants Commission (India) for providing the financial assistance in terms of Minor Research Project.

# REFERENCES

[1]   W –M Yao *et al*. (Review of Particle Properties), J. Phys. **G 33**, 1 (2006).

[2]   see for a recent review, G L Fogli, E Lisi, A Marrone and A Palazzo, Prog. Part. Nucl. Phys. **57**, 742 (2006).

[3]   see for reviews and references J L Hewett and T G Rizzo, Phys. Rep. **183**, 193 (1989); J Maalampi and M Roos, Phys. Rep. **186**, 53 (1990); A Djouadi, J Ng and T G Rizzo, in *Electroweak symmetry breaking and beyond the standard model* edited by T Barklow, S Dawson, H E Haber and S Siegrist (World Scientific, 1997) p. 416; P H Frampton, P Q Hung and M Sher, Phys. Rep. **330**, 263 (2000).

[4]   P. Langacker and D. London, Phys. Rev. **D38**, 886 (1988).

[5]   M Gronau, C N Leung and J L Rosner, Phys. Rev. **D29**, 2539 (1984); M Dittmer, A Santamaria, M C Gonzalez Garcia and J W F Valle, Nucl. Phys. **B332**, 1 (1990); A. Djouadi, Z. Phys. **C63**, 317 (1994); G Azuelos and A Djouadi, Z. Phys. **C63**, 327 (1994); F M L Almeida Jr., J H Hopes, J A Martins Simoes, P P Queiroz Filho and A J Ramalho, Phys. Rev. **D51**, 5990 (1995); J Gluza and M Zralek, Phys. Rev. **D55**, 7030 (1997); K Zuber, Phys. Rept. **305**, 295 (1998); J E Cieza Montalvo, Phys. Rev. **D59**, 095007 (1999); G Cvetic C S Kim and C W Kim, Phys. Rev. Lett. **82**, 4271 (1999); W Rodejohann and K Zuber, Phys. Rev. **D62**, 094017 (2000); F.M L Almeida Jr., Y A Coutinho, J A Martins Simoes and M A B do Vale, Phys. Rev. **D62**, 075004 (2000); *ibid*., **D63**, 075005 (2001); *ibid*., **D65**, 115110 (2002); Eur. Phys. **C22**, 277 (2001); Y A Coutinho, J A Martins Simoes and C M Porto, Eur.Phys.J. **C18**, 779 (2001); J K Singhal, S Singh, A K Nagawat and N K Sharma, Phys. Rev. **D63**, 017302 (2001); J Gluza, Acta. Phys. Pol. **B33**, 1735 (2002); F del Aguila and J A Aguilae-Saavedra, JHEP **0505**, 026 (2005); F del Aguila and J A Aguilae-Saavedra, A Martinez de la Ossa and D Meloni, Phys. Lett. **B613**, 170 (2005); S Bray, J S Lee and A Pilaftsis, Phys.Lett. **B628**, 250 (2005); F del Aguila and J A Aguilae-Saavedra and R Pittau, *Neutrino physics at large colliders*, hep-ph/0606198; T Han and B Zhang, Phys.Rev.Lett. **97**, 171804 (2006).

[6]     J. K. Singhal, S. Singh, A. K. Nagawat and N. K. Sharma, Pramana J. Phys. **59**, 465 (2002)

[7]     see, e.g., B W Lee and R E Shrock, Phys. Rev. **D16**, 1444 (1977); T P Cheng and L F Li, Phys. Rev. Lett **38**, 381 (1977); *ibid*., **45**, 1908 (1980); D Wyler and L Wolfenstein, Nucl. Phys. **B218**, 205 (1983); R N Mohapatra and J W F Valle, Phys. Rev. **D34**, 1642 (1986) J. W. F. Valle, Nucl. Phys. (Proc. Suppl.) **11,** 118 (1989); M Radial and A Santamaria, Phys. Lett. **B421**, 250 (1998); Y Okada, K Okumura and Y Shimizu, Phys. Rev. **D58**, 051901 (1998); *ibid*., **D61**, 094001 (2000); L Willmann *et al*., Phys. Rev. Lett. **82**, 49 (1999); Y Kuno and Y Okada, Rev. Mod. Phys. **73**, 151 (2001); S Lavignac, I Masina, C A Savoy, Phys. Lett. **B520**, 269 (2001); S Lavignac, I Masina and C A Savoy, Phys. Lett. **B564**, 241 (2001); S Baek, T Goto,Y Okada and K Okumura, Phys. Rev. **D64**, 095001 (2001); K A Kageyama, S Kaneko, N Shimoyama and M Tanimoto, Phys. Rev **D65**, 096010 (2002); G Cvetic, C Dib, C S Kim and J D Kim, Phys. Rev. **D66**, 034008 (2002); (E) *ibid*., **D68** 059901 (2003); *ibid*., **D71**, 113013 (2005); J Kubo, E Ma and D Suematsu, Phys. Lett. **B642**, 18 (2006).

[8]     G Cvetic, C Dib, C S Kim and J D Kim, Phys. Rev. **D74**, 093011 (2006).

[9]     G W S Hou, Nucl. Phys. Proc. Suppl. **51A,** 40 (1996).

[10]    S Y Choi, C S Kim, Y J Known and S H Lee, Phys. Rev. **D 57**, 7023 (1998).

[11]    V. Barger, S. Pakvasa and X. Tata, Phys. Lett. **B 415**, 200 (1997).

[12]    G Cvetic and C S Kim, Phys. Lett. **B 461**, 248 (1999); (E). **B 471**, 471 (2000).

[13]    F.M L Almeida Jr., Y A Coutinho, J A Martins Simoes and M A B do Vale, Phys. Lett. **B 494**, 273 (2000).

[14]    F M Renard, *Basics of Electron and Positron Collisions*, (Editions Frontiers, Gifsur Yvette, France, 1981).

[15]    K Hagiwara and D Zeppenfeld, Nucl. Phys. **B 274**, 1 (1986)

[16]    New almost-degenerate neutral heavy leptons are a feature of a number of theories of physics beyond the Standard Model, see for example, B. C.

Allanach, C. M. Harris, M.A. Parker, P. Richardson, and B. R. Webber, J. High Energy Phys. **0108**, 51 (2001).

[17]  M L Brooks *et al*. (MEGA Collaboration), Phys. Rev. Lett. **83**, 1521 (1999).

[18]  P Langacker and D. London, Phys. Rev. **D 38**, 907 (1988).

[19]  In this discussion we do not consider the mixing in charged lepton sector.

**Table.1 The explicit form of the coefficients *B* and *C*** 

| $(\lambda, \bar{\lambda})$ | $(+, 0); (0, -)$ | $(-, 0); (+, 0)$ | $(\pm, \pm)$ | $(0, 0)$ |
|---|---|---|---|---|
| $B_{\lambda, \bar{\lambda}}$ | $2\gamma$ | $2\gamma$ | $1$ | $2\gamma^2$ |
| $C_{\lambda, \bar{\lambda}}$ | $\dfrac{2(1+\beta)}{\gamma}$ | $\dfrac{2(1-\beta)}{\gamma}$ | $\dfrac{1}{\gamma^2}$ | $\dfrac{2}{\gamma^2}$ |

**Figure captions**

FIG.1: Feynman diagrams for the process $e^+ \mu^- \to W^+ W^-$ (a) light neutrino ($\nu_{\alpha L}$) exchange (b) heavy neutrino ($N_{\xi L}$) exchange in $t$- and $u$- channels.

FIG. 2: Constraints on $\left|\lambda_{e\mu}^L\right|^2$ as a function $m_\xi$.

FIG. 3: The variation of differential crass section $\dfrac{d\sigma}{d\cos\theta}$ with heavy neutrino $m_\xi$ mass for different values $\sqrt{s}$ at scattering angle $\theta = 45^0$.

FIG. 4: The variation of $\dfrac{d\sigma}{d\cos\theta}$ with $\sqrt{s}$ for different values of heavy neutrino $m_\xi$ at scattering angle $\theta = 45^0$.

FIG. 5: The variation of $\dfrac{d\sigma}{d\cos\theta}$ with $\cos\theta$ for different values of heavy neutrino $m_\xi$ at $\sqrt{s} = 200$ GeV.

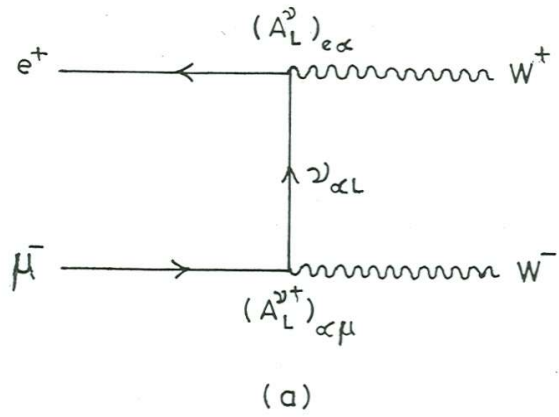

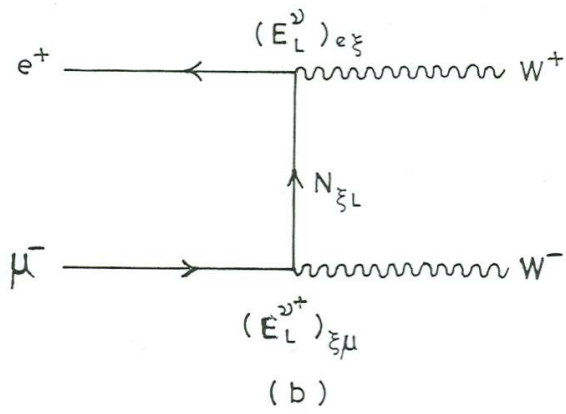

**Figure 1.** Feynman diagrams for the process $\mu^- e^+ \to W^- W^+$ (a) light neutrino ($\nu_{\alpha L}$) exchange and (b) heavy neutrino ($N_{\xi L}$) exchange.

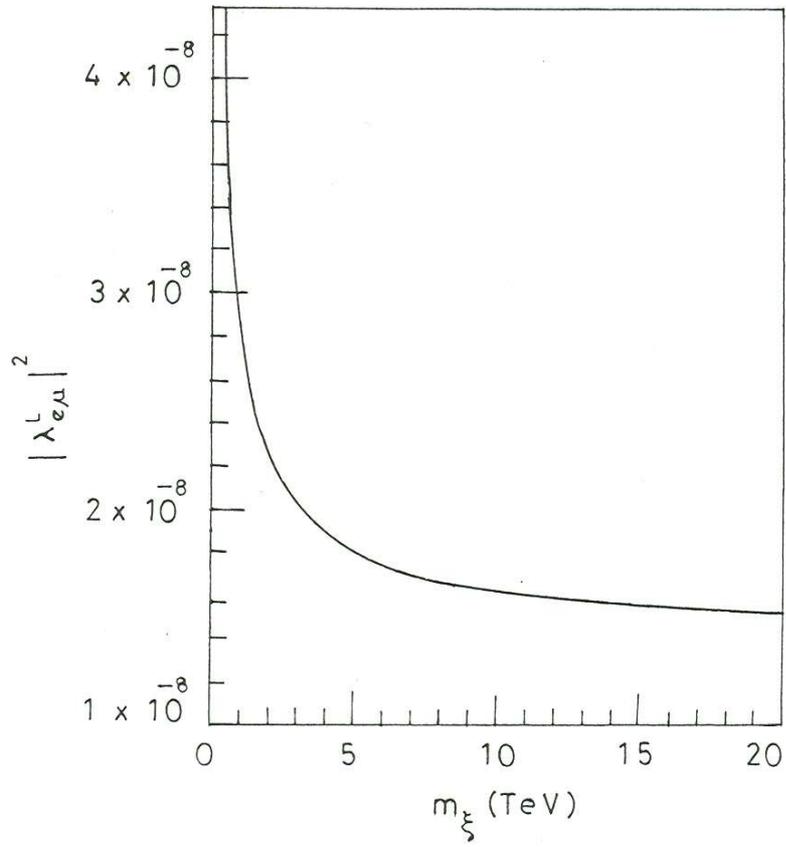

**Figure 2.** Constraints on $\left|\lambda_{e\mu}^{L}\right|^{2}$ as a function of $m_{\xi}$.

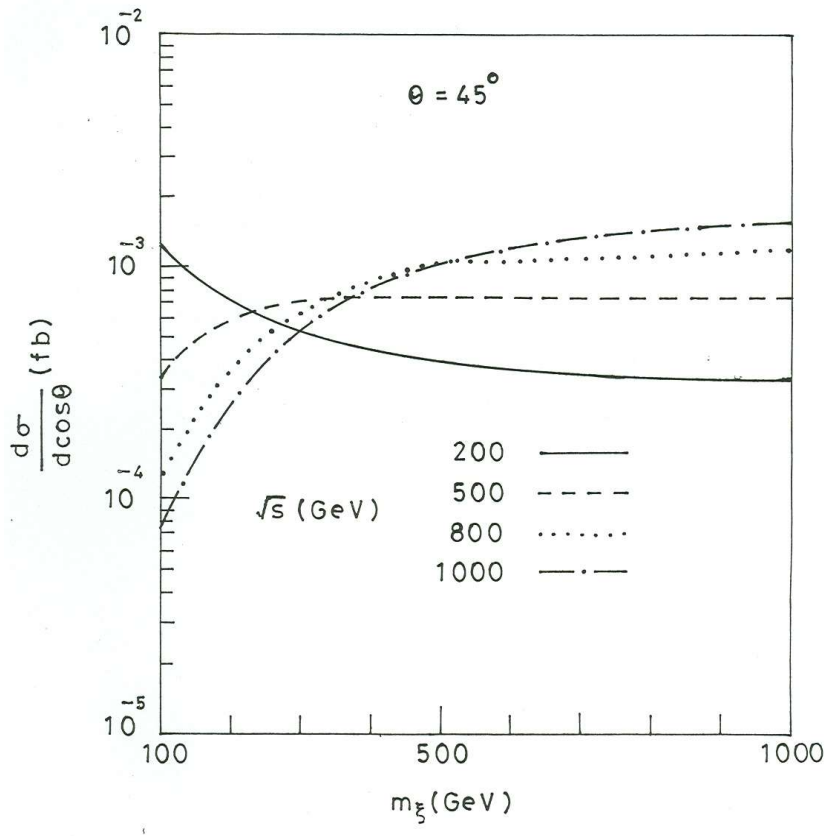

**Figure 3.** The variation of differential cross section ($\frac{d\sigma}{d\cos\theta}$) with heavy neutrino mass ($m_\xi$) for different √s at scattering angle $\theta = 45^0$ for the process $\mu^- e^+ \rightarrow W^- W^+$.

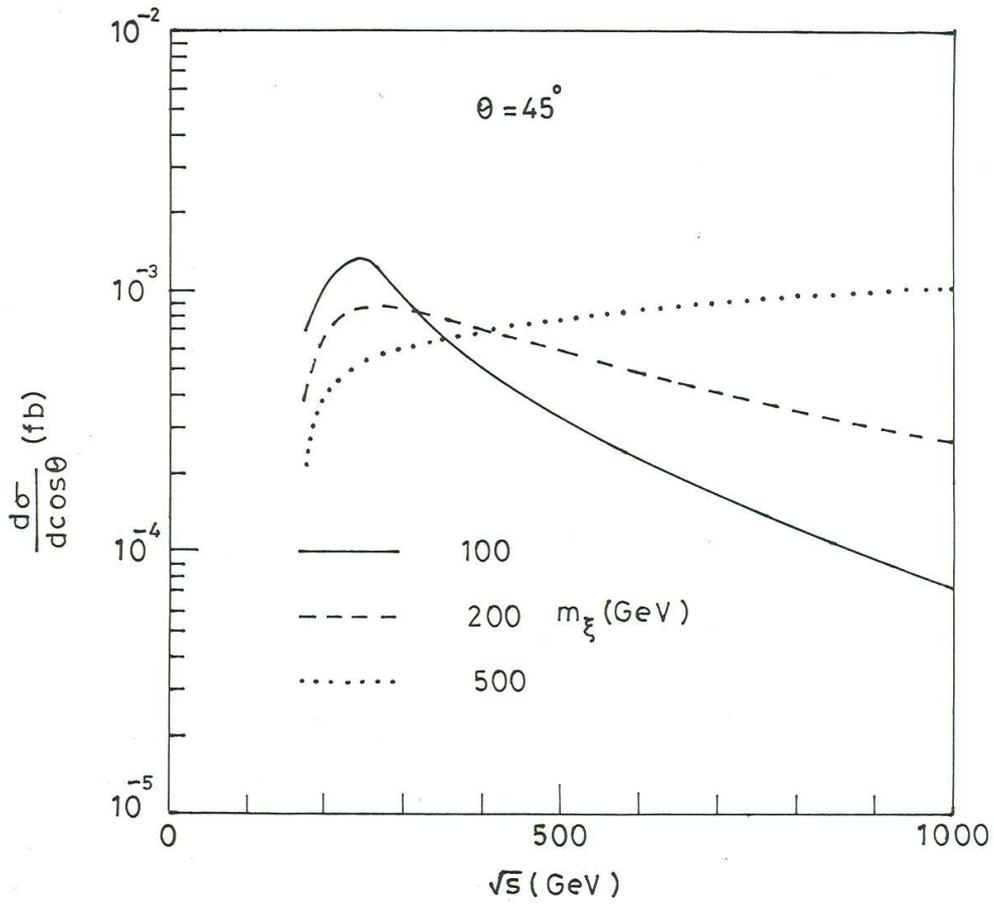

**Figure 4.** The variation of $\frac{d\sigma}{d\cos\theta}$ with $\sqrt{s}$ for different values of heavy neutrino mass ($m_\xi$) at scattering angle $\theta = 45^0$ for the process $\mu^- e^+ \to W^- W^+$.

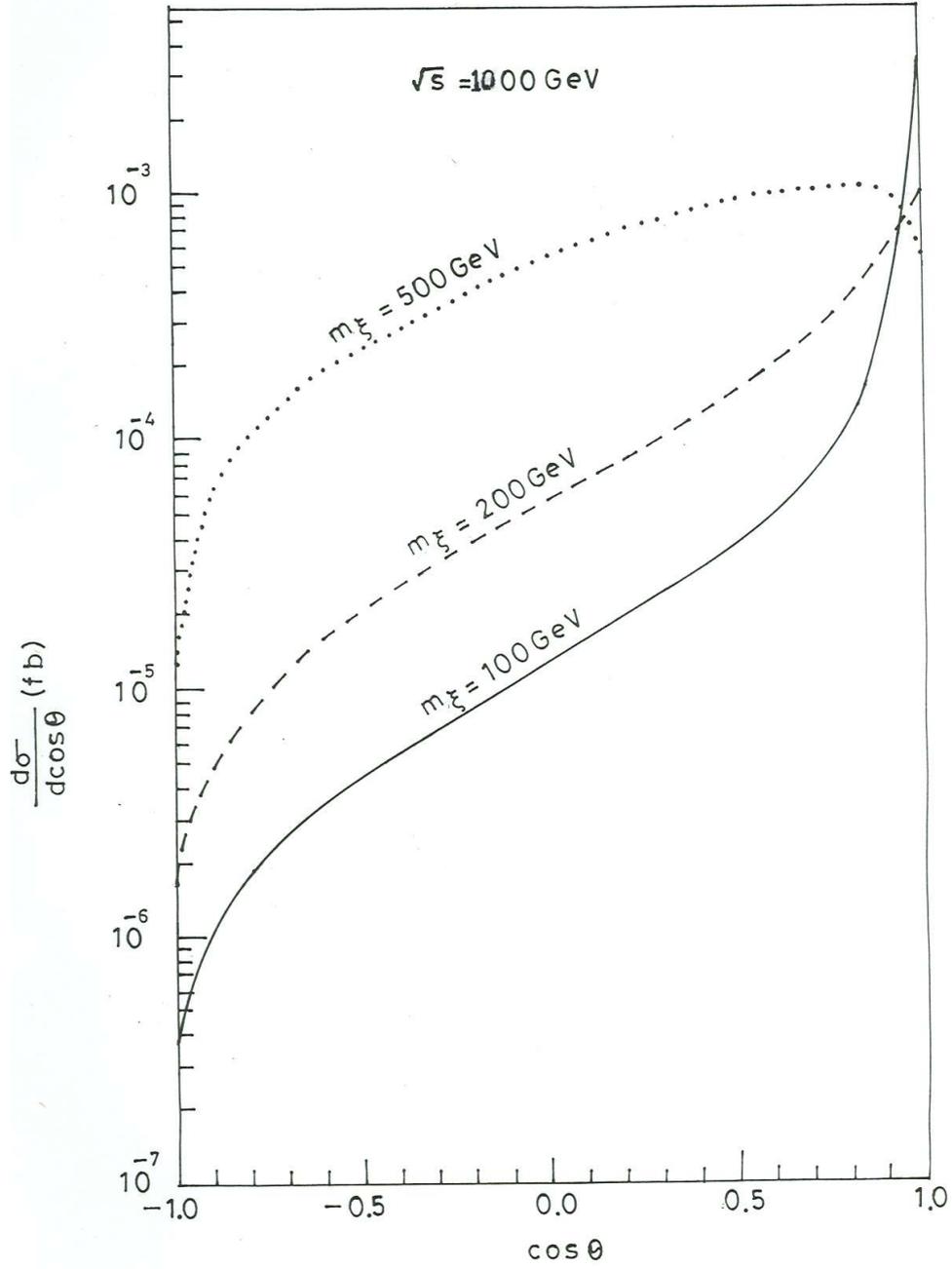

**Figure 5.** The variation of $\dfrac{d\sigma}{d\cos\theta}$ with $\cos\theta$ for different values of heavy neutrino mass ($m_\xi$) at $\sqrt{s} = 200$ GeV for the process $\mu^- e^+ \to W^- W^+$.